\documentclass[aps, prd, letterpaper, 12pt, nofootinbib, superscriptaddress, longbibliography, notitlepage]{revtex4-1}
\usepackage[utf8]{inputenc}
\usepackage{amsmath,amssymb,amsfonts}
\usepackage{mathrsfs}
\usepackage{color}
\usepackage{graphicx} 
\usepackage[section]{placeins}
\usepackage[colorlinks=true,linkcolor=blue,citecolor=blue]{hyperref}
\allowdisplaybreaks
\usepackage{orcidlink}

\pdfoutput=1 

\usepackage{graphicx}
\usepackage{hyperref}
\usepackage{aas_macros}
\usepackage{amsmath}
\usepackage{amssymb}
\usepackage{xcolor}
\usepackage{nccmath}
\usepackage{cleveref}

\def\beq{\begin{equation}\begin{aligned}}
\def\eeq{\end{aligned}\end{equation}}

\begin{document}

\title{\boldmath Primordial Black Holes and the First Stars}

\author{Julia Monika Koulen \orcidlink{0009-0001-3134-9788}}
\email{jmkoulen@ucsc.edu}
\affiliation{Department of Physics, 1156 High St., University of California Santa Cruz, Santa Cruz, CA 95064, USA}
\affiliation{Santa Cruz Institute for Particle Physics, 1156 High St., Santa Cruz, CA 95064, USA}
\affiliation{Zentrum für Astronomie und Astrophysik, Technische Universität Berlin, Hardenbergstraße 36, D-10623 Berlin, Germany}

\author{Stefano Profumo \orcidlink{0000-0002-9159-7556}}
\affiliation{Department of Physics, 1156 High St., University of California Santa Cruz, Santa Cruz, CA 95064, USA}
\affiliation{Santa Cruz Institute for Particle Physics, 1156 High St., Santa Cruz, CA 95064, USA}

\author{Nolan Smyth \orcidlink{0000-0002-8454-3015}}
\affiliation{Department of Physics, 1156 High St., University of California Santa Cruz, Santa Cruz, CA 95064, USA}
\affiliation{Santa Cruz Institute for Particle Physics, 1156 High St., Santa Cruz, CA 95064, USA}

\begin{abstract}
\noindent Primordial black holes (PBHs) constitute a compelling dark matter candidate whose gravitational effects could significantly influence early cosmic structure formation. We investigate the impact of PBHs on Population III star formation through detailed $N$-body and hydrodynamic simulations, extending beyond previous semi-analytical approaches. Our results reveal a mass-dependent dichotomy in PBH effects: massive PBHs ($M_{\rm PBH} \gtrsim 10^2 M_\odot$) with sufficient abundance can accelerate structure formation and shift Pop III formation to higher redshifts, potentially conflicting with observational constraints from high-redshift galaxy surveys. Conversely, lower-mass PBHs can induce tidal disruption of gas-rich minihalos, suppressing star formation and delaying the cosmic dawn depending on their abundance. We quantify these competing effects to derive new constraints on the PBH mass function and their contribution to the total dark matter density, with implications for forthcoming observations with the James Webb Space Telescope and 21-cm cosmology experiments.
\end{abstract}

\maketitle
\newpage
\section{Introduction}
\label{sec:intro}
Primordial black holes (PBHs)---formed from early-universe density fluctuations or from exotic late-universe processes, rather than from stellar collapse---represent a unique class of dark matter candidates  \cite{carrPrimordialBlackHoles2020a}. Unlike more diffuse conventional dark matter particles, PBHs interact gravitationally as discrete massive objects, potentially leaving distinctive imprints on cosmic structure formation and the emergence of the first stars \cite{liuImpactPrimordialBlack2024}.

Population III stars, the first generation of metal-free stars formed from primordial hydrogen and helium, represent a pivotal epoch in cosmic history. Their formation initiated stellar nucleosynthesis, cosmic reionization, and the formation of supermassive black hole seeds \cite{latifBirthMassFunction2022,osheaPopulationIIIStar2008}. In the standard $\Lambda$CDM paradigm, these primordial stars emerge within low-mass minihalos ($10^5-10^6 M_\odot$) at redshifts $z \sim 20-30$, where molecular hydrogen ($\mathrm{H}_2$) cooling enables gravitational collapse of pristine gas \cite{normanPopulationIIIStar2008}. 

The intersection of PBH physics and Pop III star formation presents a rich phenomenological landscape. PBHs can act as both catalysts and inhibitors of early star formation through competing mechanisms \cite{liuEffectsStellarmassPrimordial2022,liuImpactPrimordialBlack2024,casanueva-villarrealImpactPrimordialBlack2024}. On one hand, their gravitational influence can enhance small-scale density fluctuations and accelerate halo assembly. On the other hand, accretion-driven feedback (including heating, ionization, and Lyman-Werner radiation) can suppress $\mathrm{H}_2$ cooling and delay gravitational collapse within star-forming minihalos.

Previous investigations have employed both semi-analytical models and cosmological simulations to explore PBH effects on early structure formation \cite{liuEffectsStellarmassPrimordial2022,casanueva-villarrealImpactPrimordialBlack2024,koulenConstraintsPrimordialBlack2024}. While these studies generally conclude that PBHs exert a modest influence on Pop III formation under observationally motivated constraints, several key questions remain unresolved. First, the redshift dependence of PBH-induced modifications to the Pop III formation epoch lacks detailed quantification. Second, the transition between enhancement and suppression regimes as a function of PBH mass and abundance requires systematic exploration. Third, the implications for observational constraints on PBH dark matter fractions demand careful assessment in light of emerging high-redshift observations.

Gravitational wave detections have renewed interest in stellar-mass PBHs as dark matter constituents, while constraints from microlensing, cosmic microwave background, and big bang nucleosynthesis have shaped the viable PBH mass windows \cite{ngConstrainingHighredshiftStellarmass2022, serpicoCosmicMicrowaveBackground2020, carrPrimordialBlackHoles2022, tisserandLimitsMachoContent2007a, mrozNoMassiveBlack2024a}. Simultaneously, observations of unexpectedly massive and numerous galaxies at $z > 10$ by the James Webb Space Telescope have raised questions about whether enhanced early structure formation—potentially driven by PBHs—might reconcile theory with observations \cite{zhangHowMassivePrimordial2025}.

Our study builds upon and extends the simulation framework introduced in Ref.~\cite{liuEffectsStellarmassPrimordial2022}, which pioneered the use of cosmological zoom-in hydrodynamic simulations to investigate the influence of stellar-mass PBHs on primordial star formation. Ref.~\cite{liuEffectsStellarmassPrimordial2022} employed the GIZMO code with the meshless finite-mass (MFM) solver, non-equilibrium primordial chemistry via the GRACKLE library, and zoom-in initial conditions generated using MUSIC, focusing on the role of PBH-induced isocurvature modes and accretion feedback on early halo collapse. Our methodology adopts this powerful computational setup to ensure consistency and comparability, but diverges in scope and emphasis. Our work systematically explores the broader region of PBH parameter space, emphasizing the dichotomous role of PBHs in either enhancing or suppressing Population~III star formation. These extensions allow us to quantify the threshold between suppression and enhancement regimes, and to derive new, complementary PBH constraints.

The remainder of this study is structured as follows: Section~\ref{sec:review} reviews the theoretical framework connecting dark matter candidates to first-star formation, establishing the context for PBH-specific effects; Section~\ref{sec:sim} details our numerical simulation methodology and initial conditions; Section~\ref{sec:results} presents our findings on PBH-modified Pop III formation; Section~\ref{sec:discussion} discusses implications for dark matter constraints and observational predictions; and Section~\ref{sec:conclusions} summarizes our conclusions and future prospects.

\section{Dark Matter Candidates and Population III Star Formation}\label{sec:review}

The formation of Population III stars occurs within the broader context of cosmic structure formation driven by dark matter dynamics. While the standard $\Lambda$CDM framework successfully describes large-scale structure evolution, the microscopic nature of dark matter remains one of cosmology's most pressing questions. Different dark matter candidates can leave distinct imprints on early star formation through their unique interaction mechanisms with baryonic matter.

Two prominent dark matter candidates have received particular attention for their potential influence on Pop III formation: weakly interacting massive particles (WIMPs) and primordial black holes (see e.g. \cite{carrPrimordialBlackHoles2020a, roszkowski_wimp_2018} for recent reviews of each). These candidates represent fundamentally different physics; WIMPs are hypothetical elementary particles interacting through weak nuclear forces, and PBHs are macroscopic gravitational objects. Their contrasting properties lead to qualitatively different effects on the thermal, chemical, and dynamical evolution of star-forming gas clouds.

Understanding these interactions is crucial for several reasons. First, Pop III stars serve as sensitive probes of early universe physics, potentially encoding signatures of exotic dark matter physics in their formation redshifts, initial mass functions, and spatial distributions. Second, the relative importance of different dark matter candidates can be constrained by comparing theoretical predictions with emerging high-redshift observations \cite{houEffectsDarkMatter2025}. Finally, the interplay between dark matter physics and baryonic processes during cosmic dawn may illuminate both the nature of dark matter and the origins of cosmic structure.

This section reviews current theoretical understanding and simulation results concerning WIMP and PBH influences on Pop III star formation, establishing the foundation for our detailed numerical investigation.

\subsection{WIMP Dark Matter and Pop III Formation}

Weakly interacting massive particles represent a broad class of hypothetical dark matter candidates predicted by beyond-Standard-Model theories, including supersymmetry and extra-dimensional models. WIMPs typically possess masses in the GeV-TeV range and interact with ordinary matter through weak nuclear forces, leading to distinctive signatures in star-forming environments through two primary mechanisms: self-annihilation and elastic scattering with baryons.

The most dramatic WIMP effect on Pop III formation is the potential creation of ``dark stars''—stellar objects powered primarily by dark matter annihilation rather than nuclear fusion \cite{ioccoDarkMatterAnnihilation2008,freeseDarkMatterCapture2008,taosoDarkMatterAnnihilations2008}. This phenomenon occurs when WIMP densities in collapsing primordial halos reach critical thresholds of $\rho_\chi \sim 10^8 - 10^{12}$ GeV cm$^{-3}$, at which point annihilation heating can balance gravitational contraction and halt further collapse \cite{yoonEvolutionFirstStars2008}.

The formation mechanism proceeds through several stages. Initially, gravitational collapse of primordial gas within dark matter minihalos increases both baryonic and WIMP densities. As the central density rises, WIMP self-annihilation rates scale as $\Gamma_{\rm ann} \propto \rho_\chi^2$, rapidly increasing the energy injection rate. When annihilation heating exceeds $\mathrm{H}_2$ cooling, the proto-stellar core reaches hydrostatic equilibrium supported by dark matter energy rather than thermal pressure \cite{ioccoDarkMatterAnnihilation2008, qinBirthFirstStars2023}.

Stellar evolution calculations using modified codes such as the Geneva stellar evolution package \cite{eggenbergerGenevaStellarEvolution2008} demonstrate that dark stars can achieve remarkable properties: masses exceeding $10^3 M_\odot$, radii comparable to the present-day solar system, and main-sequence lifetimes extended by factors of $10^6-10^9$ compared to conventional Pop III stars \cite{taosoDarkMatterAnnihilations2008,yoonEvolutionFirstStars2008}. These objects exhibit distinctive observational signatures, including anomalously cool effective temperatures ($T_{\rm eff} \sim 10^4$ K) and extended evolutionary phases that could potentially be detected by next-generation telescopes.

\subsection{Primordial Black Holes and Early Star Formation}

 Primordial black holes represent a fundamentally different class of dark matter candidate. Unlike particle dark matter, PBHs interact solely via gravity and through electromagnetic processes tied to accretion~\cite{carrPrimordialBlackHoles2020a, greenPrimordialBlackHoles2021}. Recent detailed hydrodynamic simulations and semi-analytical models~\cite{liuImpactPrimordialBlack2024,zhangHowMassivePrimordial2025} show that PBH impacts on Pop~III formation depend sensitively on their mass spectrum, spatial distribution, and abundance: massive PBHs can accelerate structure formation, while lower-mass PBHs may suppress it via tidal disruption. Semi-analytical modelling in~\cite{inmanEarlyStructureFormation2019} further illuminates how PBH-induced radiation feedback shapes mini-halo evolution in the cosmic dawn.

\subsubsection{Gravitational and Dynamical Effects}

The primary gravitational effects of PBHs on early structure formation operate through several mechanisms. First, PBHs act as discrete mass concentrations that can seed enhanced density fluctuations on small scales, potentially accelerating halo formation and modifying the cosmic star formation history \cite{liuEffectsStellarmassPrimordial2022}. This effect is particularly pronounced for PBH masses in the stellar range ($M_{\rm PBH} \sim 10-100 M_\odot$), where individual objects can significantly influence the dynamics of primordial minihalos.

Second, PBH clustering and merging can create local overdensities that serve as preferential sites for early star formation. $N$-body simulations demonstrate that regions with enhanced PBH concentrations exhibit earlier halo assembly times and modified mass functions compared to standard $\Lambda$CDM predictions \cite{inmanEarlyStructureFormation2019}.

Third, PBHs can also \textit{disrupt} star formation through tidal interactions and dynamical heating. When PBH velocities exceed the virial velocities of host minihalos, gravitational encounters can strip gas from star-forming regions or prevent efficient cooling through turbulent heating \cite{casanueva-villarrealImpactPrimordialBlack2024}.

\subsubsection{Accretion Feedback and Radiative Processes}

Beyond pure gravitational effects, PBHs can influence their surroundings through accretion-driven feedback processes. When embedded within gas-rich primordial halos, PBHs accrete surrounding material at rates determined by the Bondi-Hoyle prescription (for original foundational work see Ref.~\cite{hoyleAccretionTheoryStellar1941,bondiMechanismAccretionStars1944, bondiSphericallySymmetricalAccretion1952}; for modern follow-ups see \cite{edgarReviewBondiHoyleLyttletonAccretion2004, comerfordBondiHoyleLyttletonAccretionBinary2019, blakelyRelativisticBondiHoyleLyttletonAccretion2015, cruz-osorioRelativisticBondiHoyleLyttletonAccretion2017}):
\begin{equation}
\dot{M}_{\rm acc} = 4\pi \lambda \rho_{\rm gas} \left(\frac{GM_{\rm PBH}}{c_s^2}\right)^2 c_s
\end{equation}
where $\lambda$ is a dimensionless accretion efficiency factor, $\rho_{\rm gas}$ is the local gas density, and $c_s$ is the sound speed.

This accretion process generates both thermal and ionizing radiation that can significantly impact the local intergalactic medium. Key feedback mechanisms include: \textbf{Thermal heating} from accretion luminosity heats surrounding gas, raising the minimum halo mass required for efficient $\mathrm{H}_2$ cooling and star formation; \textbf{Photoionization} from high-energy photons creates ionized regions that suppress molecular hydrogen formation through enhanced electron abundances; \textbf{Lyman-Werner radiation}---UV photons in the 11.2-13.6 eV range---dissociates $\mathrm{H}_2$ molecules, eliminating the primary coolant for primordial gas collapse.

\subsubsection{Simulation Constraints and Current Understanding}

Recent cosmological zoom-in simulations have provided quantitative constraints on PBH effects during the Pop III epoch \cite{liuEffectsStellarmassPrimordial2022,liuImpactPrimordialBlack2024,casanueva-villarrealImpactPrimordialBlack2024}. These studies generally find that, although PBHs can modify gas properties and halo assembly histories, their direct impact on individual star-forming cores is limited under observationally motivated abundance constraints.

Specifically, simulations demonstrate that PBH feedback typically increases the critical halo mass for star formation by factors of 2-5, corresponding to modest delays in Pop III formation redshifts. However, the net effect on cosmic star formation rates and the overall Pop III population appears small compared to uncertainties in conventional feedback processes such as stellar winds and supernovae \cite{koulenConstraintsPrimordialBlack2024}.

\section{Simulation Methods}
\label{sec:sim}

We investigate PBH effects on Population III star formation using high-resolution cosmological hydrodynamic simulations. Our approach employs a two-stage strategy. First, we conduct lower-resolution parent simulations to identify suitable star-forming regions. Then, we perform zoom-in simulations with enhanced resolution to capture the detailed physics of Pop III formation.

\subsection{Numerical Code and Computational Setup}

For the cosmological $N$-body simulations of Population III star formation, we employ the \texttt{GIZMO} package \cite{hopkinsGIZMONewClass2015}. \texttt{GIZMO} is a code for cosmological $N$-body simulations of structure formation using smoothed particle hydrodynamics. \texttt{GIZMO} is derived from the \texttt{GADGET} code \cite{springelCosmologicalSimulationCode2005}; it uses the parallelization scheme and Tree + PM gravity solver from \texttt{GADGET-3}, but additionally applies a Lagrangian meshless finite-mass (MFM) hydro solver.

The MFM method is an adaptive mesh-based method in which the system adapts to the matter density; it does not require a fixed regular mesh but rather kernels to compute the interactions between particles. This allows gas, stars, DM, and other astronomical objects to be simulated on a large scale.

We run the simulations on the High-Performance Computing Cluster of the Math Institute of Technische Universität Berlin. We run \texttt{GIZMO} in parallel, using between 8 and 64 CPUs depending on the particle number. The number of particles is decisive for the duration of the simulation (as the runtime of $N$-body simulations in which $N$ particles interact through gravitational force scales with $\mathcal{O}(N \log N)$). Thus, the simulations for a low number (i.e.~1) of PBH particles take about a few hours and those for a high number (i.e.~$\sim 4 \cdot 10^6$) of PBH particles up to eight days.

\subsection{Cosmological Parameters and Coordinate System}

We perform the simulations using the following cosmological parameters to implement comoving integration in the simulation: $\Omega_{\textrm{m}} = 0.3089$, $\Omega_{\Lambda} = 1 - \Omega_{\textrm{m}} = 0.6911$, $\Omega_{\textrm{b}} = 0.04864$, $n_{\textrm{s}} = 0.96$, and $h = 0.6774$ \cite{collaborationPlanck2018Results2020} corresponding to the $\Lambda$CDM model.

\subsection{Parent Simulation Strategy}

The procedure for our cosmological simulations is as follows: First, we run a parent simulation. This is essentially a pre-flight simulation in which a lower particle resolution is used to determine later areas of focus for a zoom-in simulation, which has increased resolution. This allows us to work computationally more efficiently, as we only have to focus on a very small area of the Universe during star formation and can therefore exclude surrounding regions. 

The parent simulation is a CDM-only simulation including gas and DM particles exclusively without any PBHs and follows the $\Lambda\textrm{CDM}$ model according to the mentioned cosmological parameters. We call the DM particles \textit{background DM} particles, even if no PBHs are introduced at this time. This approach allows us to study the early stages and development of the Universe which is essential to find a region of interest for simulating Pop III star formation.

\subsection{Initial Conditions and Box Setup}

We use the code \texttt{MUSIC} (Multi-scale initial conditions for cosmological simulations) to create initial conditions for the parent simulation \cite{hahnMultiscaleInitialConditions2011}. \texttt{MUSIC} allows the generation of cosmological initial conditions for a hierarchical set of nested regions, which is crucial to achieve a higher resolution for the region of interest. The initial conditions are generated at a redshift of $z = 300$. 

The simulation box is set to a size of $L \sim 200$ kpc and contains a total of $128^3$ background DM particles and $128^3$ gas particles representing the Universe during an early stage before any structures have formed. The parameter for the amplitude of density fluctuations in the cosmic microwave background (CMB) is increased to $\sigma_8 = 2$ to accelerate structure formation. As shown in \cite{parkFirstStructureFormation2020}, increasing $\sigma_8$ does not affect the Population III star formation rate for a given halo mass, as the relation between halo mass and star formation remains unchanged.
During the parent simulation, the masses of the particles are set to $m_{\textrm{BDM}} \sim 140 ~M_{\odot}$ and $m_{\textrm{gas}} \sim 26 ~M_{\odot}$. The choice of the values for the particle number and the associated particle masses is based on the approach in \cite{liuEffectsStellarmassPrimordial2022}. 

\subsection{Gravitational Softening}

In $N$-body simulations, the softening length $\epsilon$ is essential for mitigating numerical issues that arise on small scales. When the distance $r_{12}$ between two particles with masses $m_1$ and $m_2$ approaches zero, the gravitational force follows a $\sim 1/r_{12}^2$ dependence, which can lead to arbitrarily large accelerations and numerical instabilities. In physical systems, such effects would be naturally regulated by collisional interactions and the finite size of the masses. However, since particles in the simulation are treated as point masses, the gravitational force must be softened as $r_{12} \rightarrow 0$ to prevent artificial divergences. Therefore, the softening length is introduced, as described in \cite{rodionovOptimalChoiceSoftening2005}.

\begin{equation}
    F_{\textrm{soft}} = \frac{G m_1 m_2}{r_{12}^2 + \epsilon^2}.
\end{equation}

For $r\gg\epsilon$ the expression $F_{\textrm{soft}}$ approaches the regular gravitational force $F$ and for $r\ll\epsilon$ the gravitational force approaches a maximal value for the two interacting masses. $\epsilon$ thus determines the distance below which the gravitational force between two particles is reduced, ensuring that it does not become infinitely large as the separation between particles approaches zero. The choice of an appropriate softening length is an important aspect of $N$-body simulations as it has a significant impact on the accuracy of the simulated astrophysical systems \cite{adamekWIMPsStellarmassPrimordial2019}.

For the softening lengths of the gas and background DM (BDM) particles, the comoving softening length of $\epsilon_{\text{BDM}} = \epsilon_{\text{gas}} = 0.01 h^{-1}$ kpc is used for parent and zoom-in simulation, which corresponds to the values of the softening lengths used in \cite{liuEffectsStellarmassPrimordial2022}.

\subsection{Primordial Chemistry and Cooling Implementation}

To achieve Pop III star formation in a cosmological $N$-body simulation, cooling mechanisms must be implemented into the code which are based on a non-equilibrium chemistry. This allows for the representation of cooling processes within a primordial gas at very high redshifts. To achieve this, the external cooling library \texttt{GRACKLE} is integrated into the code. \texttt{GRACKLE} provides the option for cooling based on a non-equilibrium primordial chemistry network, enabling the construction of a 12-species network, including $\mathrm{H}, \mathrm{H}^+, \mathrm{H}^-, \mathrm{e}^-, \mathrm{He}, \mathrm{He}^+$, $\mathrm{He}^{2+}, \mathrm{H}_2, \mathrm{H}_2^+, \mathrm{D}, \mathrm{D}^+$, and $\mathrm{HD}$ \cite{smithGrackleChemistryCooling2017}.

\subsection{Simulation Evolution and Termination Criteria}

During the cooling process, typical star-forming DM minihalos emerge. In this phase, gas particles start accreting, which causes the accumulation of gas within the evolving halos. The simulation is stopped when the densest gas particle reaches a critical hydrogen number density value of $n_{\textrm{H}} \gtrsim 10^4$ cm$^{-3}$, representing a standard criterion for the beginning of Pop III star formation. At this point, gas clouds at the halos' centers have formed as a result of accretion processes induced by cooling. This stage, characterized by gas clouds entering runaway-collapse with $n_{\textrm{H}} \gtrsim 10^2$ cm$^{-3}$ and $T \lesssim 500$ K \cite{liuEffectsStellarmassPrimordial2022}, denotes the transition from the initial scattered distribution of DM and gas particles to the formation of the first structures. We take this as a termination point for our simulations since the criterion represents a typical condition for Pop III star formation.

\subsection{Halo Finding and Zoom-in Region Selection}

After the parent simulation is terminated, the \texttt{ROCKSTAR} halo finder \cite{behrooziROCKSTARPhasespaceTemporal2013} is used to identify DM halos in the simulation box. The halo finder is based on the friends-of-friends algorithm, which identifies halos by considering particles that are within a certain distance of each other. If a particle is closer to another particle than the linking length $l$, they are considered as \textit{friends}. Additionally, all particles linked to the friends of friends are included in the group, which is classified as a halo.

We identify a particular DM minihalo with a virial mass of $M_{\text{halo}} \sim 1.2 \times 10^6 M_{\odot}$ and a virial radius of $r_{\text{vir}} \sim 111$ pc at a redshift of $z \sim 29$. For the zoom-in simulation, the resolution in the region of interest is increased by two levels, i.e., $\Delta_{\text{res}} = 2$. For this purpose, a Lagrangian region is defined as $R_{\text{L}} = (1.5 \Delta_{\text{res}} + 1)R_{\text{vir}} = 4R_{\text{vir}}$ \cite{onorbeHowZoomBias2014}. This approach prevents the contamination of the region of interest by low-resolution particles.

\subsection{Zoom-in Simulation Configuration}

All background DM and gas particles located within the Lagrangian region are now identified using their particle IDs and traced back to their initial positions at $z = 300$. A box, limiting the high-resolution region, is placed around these particles. After identifying the zoom-in box, the initial conditions can be generated for the zoom-in simulation. In addition to the coarse grid of the parent simulation, they include a refinement region with two additional levels for the cell grid. This increases the particle number in this area, resulting in lowered masses for background DM and gas particles which are now given by $m_{\text{BDM}} \sim 2.17 ~M_{\odot}$ and $m_{\text{gas}} \sim 0.4 ~M_{\odot}$. Particles outside the zoom-in region are described by coarse-level particles no longer representing the types of background DM and gas particles.

\subsection{Comparison with Previous Work}

Our simulation pipeline is closely aligned with the framework developed in Ref.~\cite{liuEffectsStellarmassPrimordial2022}, which first combined cosmological zoom-in simulations with PBH physics to study early structure formation. We employ the same gravitational and hydrodynamic solvers (GIZMO with MFM and TreePM gravity), non-equilibrium cooling chemistry (GRACKLE), initial condition generation (MUSIC), and halo identification (ROCKSTAR). This deliberate consistency allows us to isolate physical effects due to PBH mass and abundance, rather than differences in numerical implementation.

Despite the shared foundation, our study differs in several critical aspects:
\begin{itemize}
    \item We explore the 2-D parameter space of PBH masses \textit{and} abundances, enabling a systematic mapping of the boundary between enhancement and suppression of Pop III star formation, as well as the determination of PBH constraints. 
    \item We emphasize the mass-dependent \emph{bifurcation} in PBH effects—tidal suppression versus early seeding—rather than focusing solely on feedback from accretion.
    \item We perform multiple realizations (10 per parameter point) to assess stochastic variations in collapse redshift and quantify uncertainties.
\end{itemize}
These novel contributions outside the scope of Ref.~\cite{liuEffectsStellarmassPrimordial2022}, complement and extend that work.

\section{Simulation Results}
\label{sec:results}

\subsection{CDM Simulation}
\label{popIII:formation}

The zoom-in simulation is terminated when the maximum hydrogen number density—the highest value among all gas particles in the dark matter (DM) minihalo—exceeds $n_{\textrm{H}} \gtrsim 10^5 \, \textrm{cm}^{-3}$. This threshold defines the collapse redshift $z_{\text{col}}$. This criterion ensures that a gas clump has formed in which several gas particles surpass the threshold of $n_{\textrm{H}} \gtrsim 10^4 \, \textrm{cm}^{-3}$, corresponding to the standard criterion for Pop III star formation. Consequently, we extend the zoom-in simulation beyond the parent simulation, which was terminated when the densest gas particle first reached the critical $n_{\textrm{H}}$ value. As a result, the redshift and halo mass at termination differ between the zoom-in and parent simulations. Note that this simulation excludes PBHs, representing the standard $\Lambda$CDM scenario.

\begin{figure}
\centering
\includegraphics[width=0.9\textwidth]{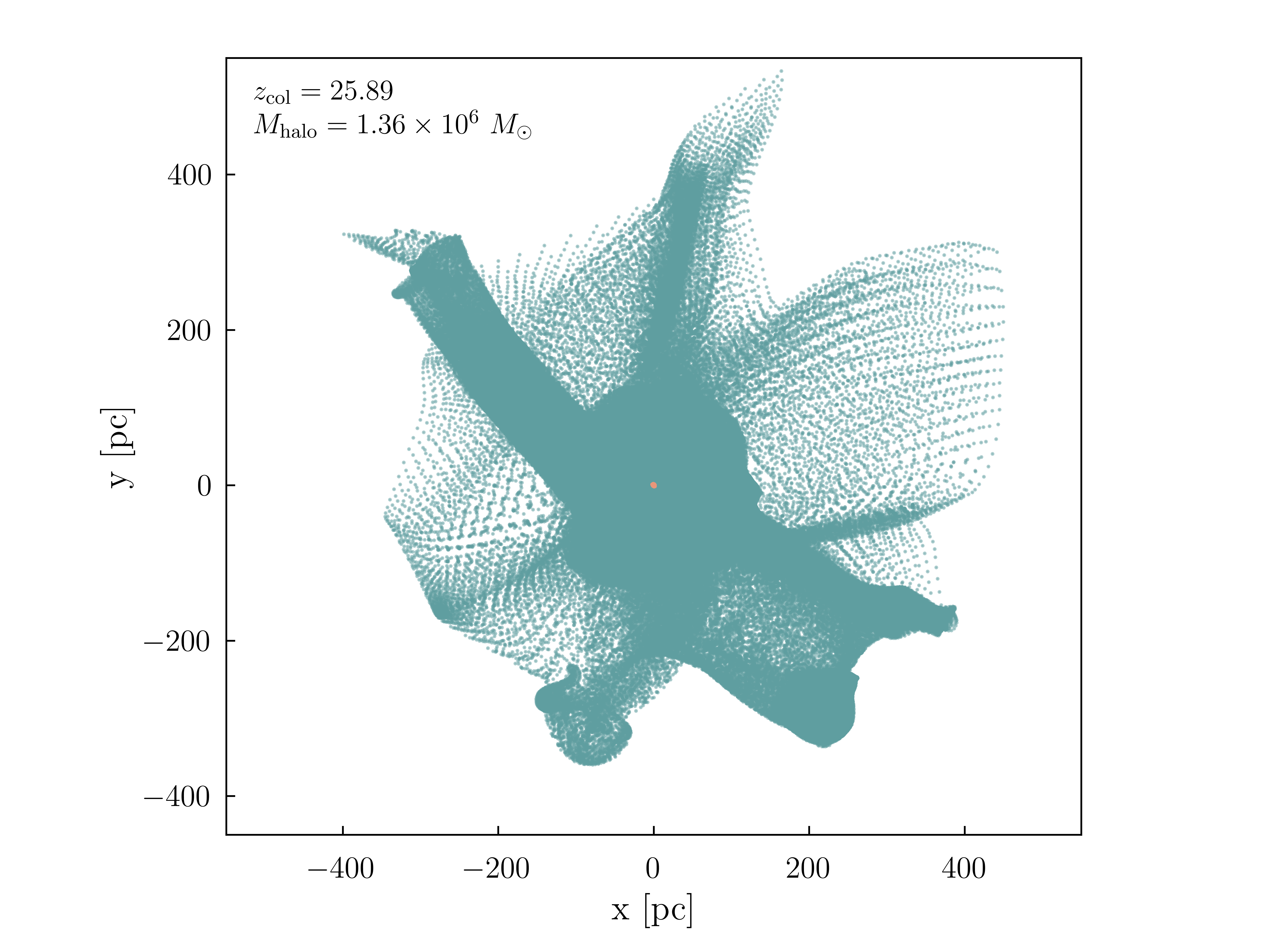}
\caption{\textbf{Spatial distribution of particles in the zoom-in CDM simulation at collapse redshift $z_{\text{col}} \sim 26$.} The projection onto the $xy$-plane shows blue background DM particles and orange gas particles with $n_{\textrm{H}} \gtrsim 10^4 \, \textrm{cm}^{-3}$. The gas particles form a coherent clump with mass $M_{\text{clump}} \sim 4.8 \times 10^3 \, M_{\odot}$ and extent $r \sim 1.2 \, \text{pc}$. This pure CDM simulation contains no PBHs. All units are in physical coordinates accounting for the cosmic scale factor. While the collapse redshift $z_{\rm col}$ is also used as a diagnostic in Ref.~\cite{liuEffectsStellarmassPrimordial2022}, here we systematically map its variation across a grid of PBH masses and abundances, thereby identifying the transition between suppression and enhancement regimes.}
\label{fig:CDM_only}
\end{figure}

Figure~\ref{fig:CDM_only} presents a snapshot of the zoom-in box at the collapse redshift $z_{\text{col}} \sim 26$. All background DM particles are shown in blue, including both halo particles and particles integrated into the high-resolution box according to the Lagrangian region definition described in Section~\ref{sec:sim}. We determine the halo center and mass using the Rockstar halo finder, enabling calculation of the virial radius. The black circle indicates the DM halo's virial radius of $r_{\text{vir}} \sim 146 \, \text{pc}$, while the halo mass at collapse is $M_{\text{halo}} = 1.36 \times 10^6 \, M_{\odot}$. Orange particles represent gas particles with hydrogen number densities exceeding $n_{\textrm{H}} \gtrsim 10^4 \, \textrm{cm}^{-3}$ at $z_{\text{col}} \sim 26$.

Figure~\ref{fig:CDM_only} demonstrates the formation of a DM minihalo where cooling mechanisms trigger runaway collapse, resulting in a central gas clump with density exceeding $n_{\textrm{H}} \gtrsim 10^4 \, \textrm{cm}^{-3}$. This gas clump, with mass $M_{\text{clump}} \sim 4.8 \times 10^3 \, M_{\odot}$ and extent $r \sim 1.2 \, \text{pc}$, satisfies the standard criterion for Pop III star formation and represents an early stage of Pop III stellar evolution. Table~\ref{tab:zoom-in_halo} summarizes the characteristic masses of the target halo.

\begin{table}
    \centering
    \begin{tabular}{|c|c|}
        \hline
        Quantity & Value \\
        \hline
        Virial mass of the halo $M_{\text{halo}}$ &  $1.36 \times 10^6 \, M_{\odot}$ \\
        Mass of a gas particle $m_{\textrm{gas}}$ & $0.4 \, M_{\odot}$ \\
        Mass of a DM particle $m_{\textrm{BDM}}$ & $2.1 \, M_{\odot}$ \\
        Total gas mass $M_{\textrm{gas}}$ & $2.18 \times 10^5 \, M_{\odot}$\\
        Total DM mass $M_{\textrm{DM}}$ & $1.14 \times 10^6 \, M_{\odot}$ \\
        \hline
    \end{tabular}
    \vspace{12pt}
    \caption{\textbf{Mass quantities of the target DM halo.}}
    \label{tab:zoom-in_halo}
\end{table}

\subsection{Including PBHs}
\label{popIII:including_PBHs}

Having established Pop III star formation in the CDM case, we now incorporate PBHs into the zoom-in simulation initial conditions to investigate their influence on early star formation processes described in Section~\ref{popIII:formation}.

We inject PBHs following the phase-space distribution of background DM particles in the zoom-in simulation. This six-dimensional phase space encompasses three position coordinates $(x, y, z)$ and three velocity components $(v_{\text{x}}, v_{\text{y}}, v_{\text{z}})$. We learn this joint distribution using a masked autoregressive normalizing flow, which maps a simple Gaussian distribution to the observed distribution through invertible transformations \cite{durkanNeuralSplineFlows2019, papamakariosMaskedAutoregressiveFlow2018}. Once trained, the flow generates new samples from the learned distribution with high computational efficiency. We validate this approach by comparing unseen test data with generated samples in both observable phase space and learned latent space, confirming distributional similarity between original and generated samples.

To maximize gravitational effects and comprehensively understand PBH impact on Pop III star formation, we consider a broad range of PBH mass $m_{\textrm{PBH}}$ and abundance $f_{\textrm{PBH}}$. We adopt a monochromatic mass function, ensuring all PBHs within a given scenario share identical mass $m_{\textrm{PBH}}$. We vary the PBH abundance $f_{\text{PBH}}$ by adjusting the number of PBHs, $n_{\text{PBH}}$. The total DM halo mass $M_{\textrm{DM}}$ includes both existing background DM particles from the CDM case (Subsection~\ref{popIII:formation}) and injected PBHs. To maintain constant $M_{\textrm{DM}}$, we proportionally reduce the individual background DM particle mass $m_{\textrm{BDM}}$ (Table~\ref{tab:zoom-in_halo}) when introducing PBHs of varying masses.

To account for stochastic variations in initial conditions, we perform 10 independent runs for each $(m_{\textrm{PBH}}, f_{\textrm{PBH}})$ combination, varying only PBH positions and velocities while maintaining fixed gas and background DM particle configurations. This approach preserves the original configuration that enables Pop III star formation in the target halo. We adopt a comoving PBH softening length of $\epsilon_{\text{PBH}} = 10^{-3} \, h^{-1} \, \text{pc}$ following \cite{liuEffectsStellarmassPrimordial2022}, which is significantly smaller than the spatial extent of the star-forming gas cloud at the DM minihalo center.

\begin{figure}
\centering
\includegraphics[width=0.75\textwidth]{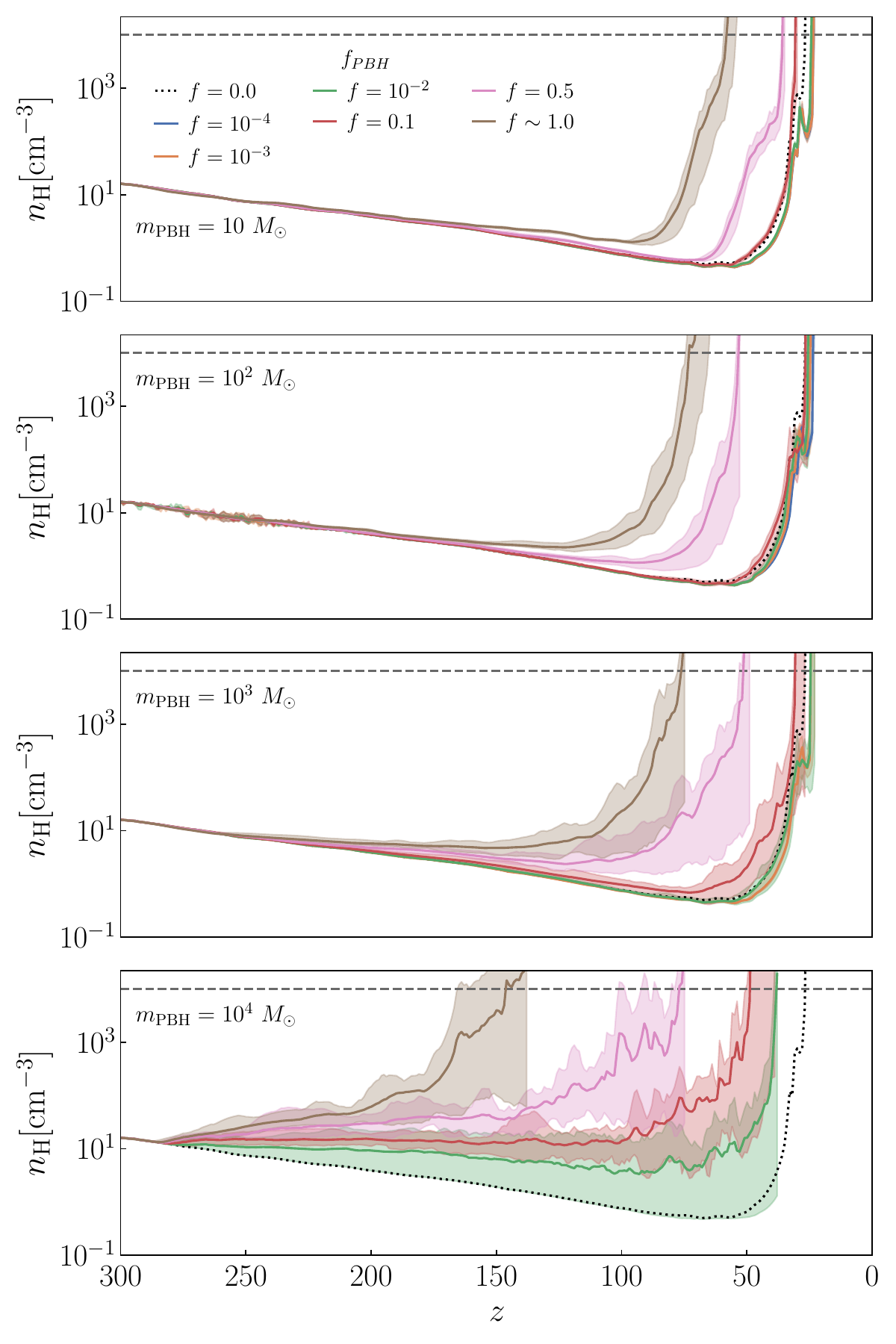}
\caption{\textbf{Redshift evolution of maximum hydrogen number density in DM halos containing PBHs.} Each panel displays the evolution of the densest gas particle's hydrogen number density for different PBH mass and abundance scenarios. Solid curves represent averages over 10 independent simulations, while shaded regions indicate the range between minimum and maximum values across all realizations. The horizontal dashed line marks the Pop III star formation threshold $n_{\mathrm{H}} = 10^4 \, \textrm{cm}^{-3}$. The intersection points reveal how PBH properties influence the timing of structure collapse and subsequent star formation.}
\label{fig:n_H_over_z_combined}
\end{figure}

Figure~\ref{fig:n_H_over_z_combined} shows the temporal evolution of the maximum hydrogen number density as a function of redshift for PBH abundances $f_{\text{PBH}} = [0 - 1.0]$ and masses $m_{\text{PBH}} = [10 - 10^4] \, M_{\odot}$. The intersection of each curve with the horizontal dashed line marks the collapse redshift where hydrogen number density exceeds the critical threshold $n_{\textrm{H}} \gtrsim 10^4 \, \textrm{cm}^{-3}$. Solid curves represent averages over 10 simulations with different PBH initial conditions, while shaded regions span the minimum and maximum values across all realizations.

Based on Figure~\ref{fig:n_H_over_z_combined}, PBH presence shifts the collapse redshift, altering structure formation timing. For very massive PBHs ($m_{\mathrm{PBH}} = 10^4 \, M_{\odot}$), their presence exclusively accelerates Pop III star formation. The maximum hydrogen number density exceeds the critical threshold at earlier redshifts for all investigated PBH abundances compared to the CDM case ($f_{\text{PBH}} = 0.0$). This premature structure formation and earlier Pop III star formation results from PBHs acting as cosmic structure generators through seed and Poisson effects \cite{carrPrimordialBlackHoles2018, meszarosBehaviourPointMasses1974}, as discussed in detail in Section~\ref{sec:discussion}.

For lower-mass PBHs, a contrasting phenomenon emerges: while very high PBH abundances still accelerate Pop III star formation relative to the PBH-free case, low abundances delay the critical maximum hydrogen number density threshold—postponing Pop III star formation onset. This delay manifests for PBH mass $m_{\mathrm{PBH}} = 10^3 \, M_{\odot}$ at abundance $f_{\mathrm{PBH}} \leq 10^{-2}$, and for smaller PBH masses at abundances $f_{\mathrm{PBH}} \leq 0.1$.

Unlike the first scenario where PBHs induce small-scale density fluctuations that accelerate Pop III formation, smaller PBH masses and fractions appear to delay the process. This delay arises from tidal effects when gas clouds encounter PBHs. The higher mass of PBHs ($m_{\mathrm{PBH}} = 10, 10^2 \, M_{\odot}$) compared to gas particles increases gas particle kinetic energy and velocities, generating heating that counteracts ongoing cooling processes. At sufficiently high PBH fractions ($f_{\mathrm{PBH}} > 10^{-2}$), the PBH population becomes large enough to delay gas cloud cooling at the DM minihalo center, ultimately postponing Pop III star formation onset beyond the PBH-free timeline.

\subsection{Constraints on PBHs}
\label{subsec:constraints}

By analyzing PBH impact on Pop III star formation, we can establish constraints on PBHs as a dark matter component. We assume a critical redshift $z_{\text{crit}}$ beyond which Pop III star formation ceases, based on the premise that Pop III stars formed only up to a specific redshift, with subsequent star formation producing exclusively Pop II and Pop I stars.

For constraint derivation, we consider different redshift thresholds: $z \sim 30, 40, 120$. These correspond to collapse redshifts resulting from the smallest possible PBH fractions for different masses, representing lower limits from minimum $f_{\text{PBH}}$ values for specific masses.

\begin{figure}
\centering
\includegraphics[width=0.9\textwidth]{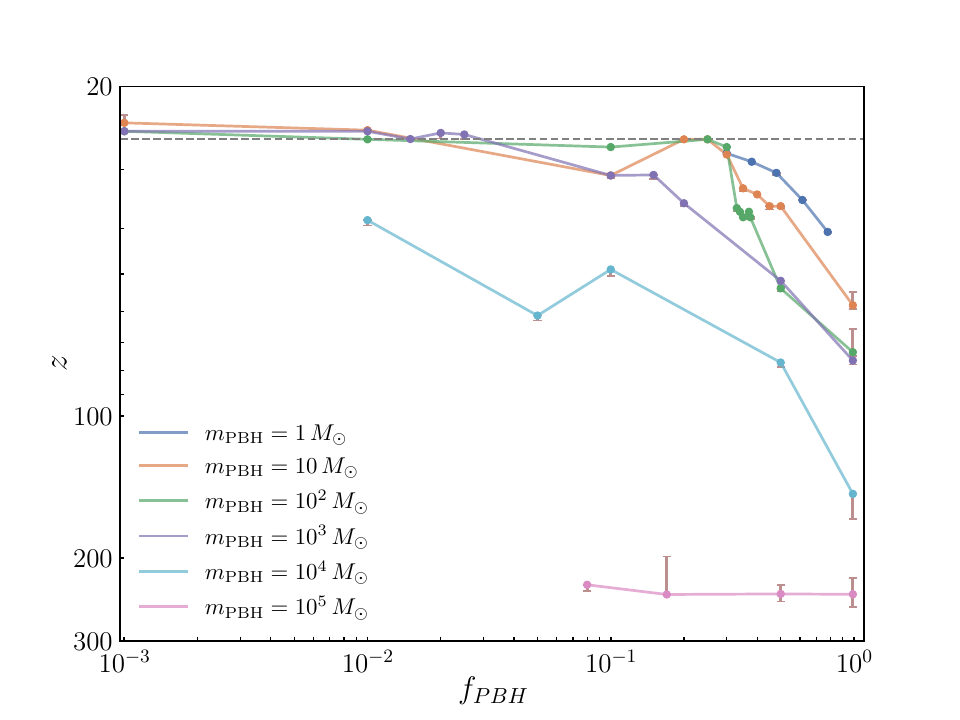}
\caption{\textbf{Critical collapse redshift as a function of PBH abundance for different PBH masses.} For each analyzed PBH mass and abundance configuration, we show the critical redshift at which maximum hydrogen number density first exceeds $n_{\textrm{H}} \gtrsim 10^4 \, \textrm{cm}^{-3}$. Intermediate fraction values are determined by interpolation between simulation points. Error bars span the range between minimum and maximum redshift values across all simulation realizations at which the density threshold is first exceeded (see Figure~\ref{fig:n_H_over_z_combined}). The horizontal gray dashed line marks the collapse redshift $z_{\text{col}} \sim 26$ for the PBH-free case ($f_{\mathrm{PBH}} = 0$).}
\label{fig:z_crit_f_PBH}
\end{figure}

Using results from Figure~\ref{fig:n_H_over_z_combined} and additional simulations for the evolution of the redshift at lower, upper, and intermediate values of the PBH mass, i.e.~$m_{\mathrm{PBH}} = 1\, M_{\odot}$ and $m_{\mathrm{PBH}} = 10^5 \, M_{\odot}$ (not shown in Figure~\ref{fig:n_H_over_z_combined}), we determine collapse redshifts for all considered PBH masses and abundances where $n_{\textrm{H}} \gtrsim 10^4 \, \textrm{cm}^{-3}$, yielding Figure~\ref{fig:z_crit_f_PBH}. To estimate the value of a PBH fraction $f_{\mathrm{PBH}}$ corresponding to the critical redshift, we apply linear interpolation between the two simulated data points that enclose the respective redshift threshold. Error bars represent minimum and maximum redshift values across the simulation ensemble, derived from the intersection of the shaded region boundaries and the $n_{\mathrm{H}} = 10^4 \, \textrm{cm}^{-3}$ threshold line in Figure~\ref{fig:n_H_over_z_combined}.

\begin{figure}
\centering
\includegraphics[width=0.8\textwidth]{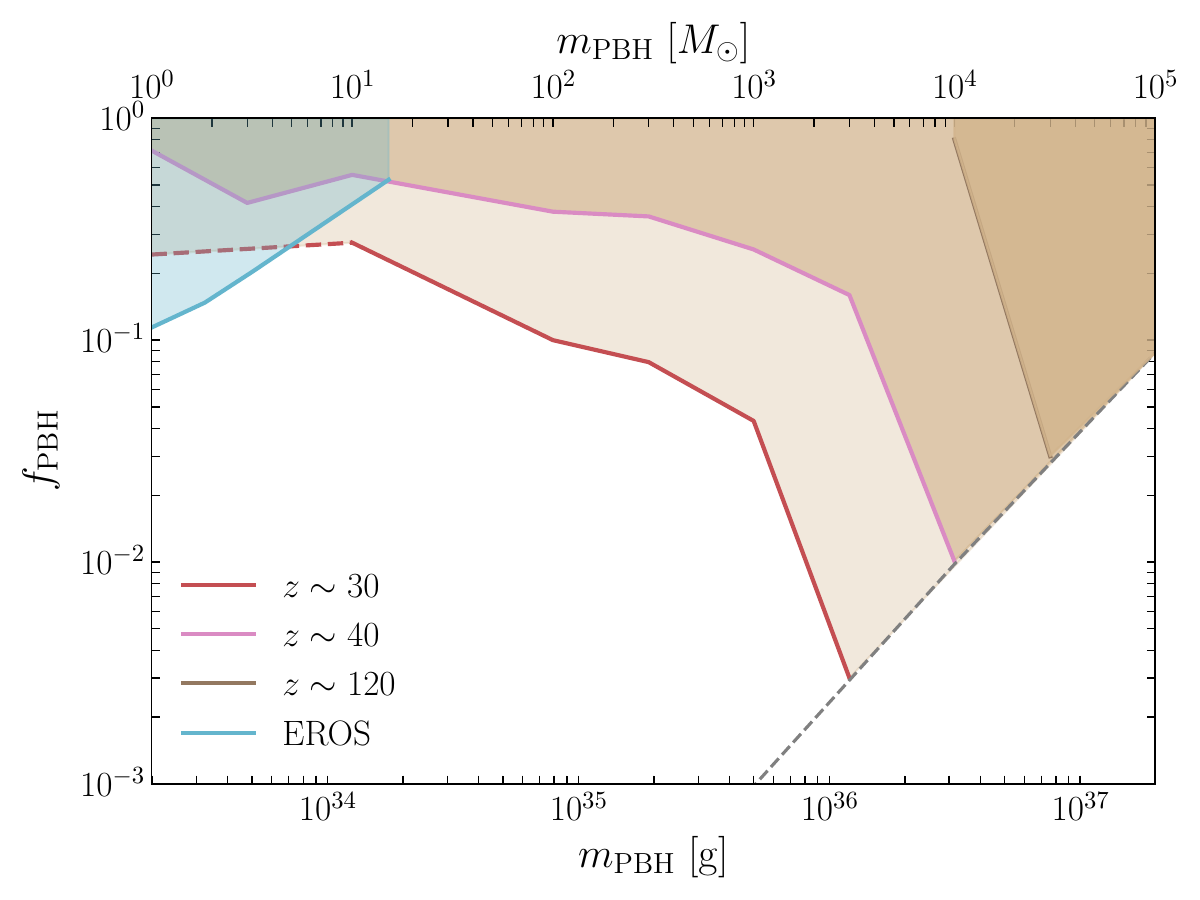}
\caption{\textbf{Constraints on PBH abundance as a function of PBH mass.} Exclusion limits are derived using average hydrogen number density evolution over 10 simulations for each mass, assuming different critical redshifts $z_{\text{crit}}$ beyond which Pop III star formation ceased. Regions above each curve are excluded for the corresponding redshift assumption. The gray dashed line indicates the physical boundary where PBH mass would exceed the total halo mass. The red dashed line shows an extrapolated constraint at $z \sim 30$ for PBH masses below $10 \, M_\odot$. These constraints demonstrate how observations of Pop III star formation timing can constrain the primordial black hole contribution to dark matter. The light blue line indicates the PBH constraints obtained from the EROS-2 microlensing observations of the Magellanic Clouds \cite{tisserandLimitsMachoContent2007a}.}
\label{fig:constraints_popIII_new}
\end{figure}

Figure~\ref{fig:constraints_popIII_new} presents constraints on $f_{\mathrm{PBH}}$ and $m_{\mathrm{PBH}}$ for redshifts $z \sim [30, 40, 120]$, derived from intersections between these fixed redshift values and curves in Figure~\ref{fig:z_crit_f_PBH}. We also performed simulations for the redshift evolution of the intermediate masses $m_{\mathrm{PBH}} = 3 \times 10^2 \, M_{\odot}$ and $m_{\mathrm{PBH}} = 3 \times 10^3 \, M_{\odot}$, and calculated the corresponding critical redshifts analogously to the other masses shown in Figure~\ref{fig:z_crit_f_PBH}. The results are included in Figure~\ref{fig:constraints_popIII_new}. To also illustrate the limiting behavior of the model at lower redshifts and PBH masses below $m_{\mathrm{PBH}} = 10 \, M_{\odot}$, we include an extrapolated constraint for $z \sim 30$, depicted by the red dashed line. The gray dashed line represents the physical boundary where a single PBH of mass $m_{\mathrm{PBH}}$ at fraction $f_{\mathrm{PBH}}$ would exceed the total halo mass, excluding unphysical parameter combinations. We also show the line, and shade in blue, corresponding to constraints from the EROS-2 microlensing observations of the Magellanic Clouds \cite{tisserandLimitsMachoContent2007a}.

\section{Discussion}
\label{sec:discussion}

Our numerical simulations demonstrate a spectrum of phenomenology based on PBH mass. High mass PBHs universally accelerate structure formation, the rate of which is determined by their abundance. As shown in Figure \ref{fig:n_H_over_z_combined}, for identical PBH abundances $f_{\text{PBH}}$ but varying numbers and masses, scenarios with more massive PBHs exhibit the most rapid hydrogen number density evolution. Consequently, massive PBHs in large numbers most significantly amplify structure formation, shifting collapse to extremely early redshifts. This enhanced structure formation is visually evident in projection snapshots where background DM particles cluster preferentially around PBHs. However, lower mass PBHs can accelerate or \textit{delay} Pop III star formation depending on their abundance.

The interplay between these regimes can be understood as follows: PBHs source density fluctuations affect surrounding matter through two distinct mechanisms. The {\bf seed effect} (also known as the Coulomb effect) originates from individual PBHs, generating initial density fluctuations of magnitude $\frac{m_{\text{PBH}}}{M}$, where $M$ represents the mass of surrounding matter influenced by the fluctuation. These seeds function as gravitational centers that can trigger cosmic structure formation. The {\bf Poisson effect}, conversely, arises when $N$ PBHs are distributed within a given region, producing statistical fluctuations in PBH number of magnitude $\sqrt{N}$. This generates initial density fluctuations of size ${(f_{\text{PBH}} \, m_{\text{PBH}}/M)}^{1/2}$ within that region. The Poisson effect is also termed Poisson noise, as it introduces discreteness noise on small scales due to the discrete nature of PBHs \cite{carrStatisticalClusteringPrimordial1977}. 

The seed effect dominates when PBH abundance is very low ($f_{\text{PBH}} \rightarrow 0$), where non-linear, gravitationally bound background DM structures form around individual PBHs. Since the PBH fraction is small, these objects do not interact with each other and can be treated as isolated seeds \cite{liuEffectsStellarmassPrimordial2022}. In the opposite limit ($f_{\text{PBH}} \rightarrow 1$), the Poisson effect dominates. At modest fractions ($f_{\rm PBH}\lesssim10^{-2}$), lower-mass PBHs delay collapse through dynamical and tidal heating, increasing gas temperature and inhibiting molecular cooling. Only at higher abundances ($f_{\rm PBH}\gtrsim10^{-1}$) do growth effects override this suppression.

In cases where the PBH mass is close to the mass of background dark matter (DM) particles, numerical artifacts may arise due to discreteness effects and unphysical two-body interactions. In our zoom-in simulations, the mass of a high-resolution background DM particle is \( m_{\rm BDM} \approx 2.1\,M_\odot \), implying a mass ratio \( M_{\rm PBH} / m_{\rm BDM} \sim 5 \) for \( M_{\rm PBH} = 10\,M_\odot \). Such a small ratio may lead to enhanced stochastic gravitational scattering between PBHs and DM particles, potentially introducing spurious heating, diffusion, or artificial dynamical friction.

We mitigate these issues through several techniques. First, we employ a small comoving gravitational softening length for PBHs (\( \epsilon_{\rm PBH} = 0.01\,h^{-1}\,\mathrm{kpc} \)), which suppresses short-range forces and reduces hard encounters. Second, each simulation is repeated over 10 independent Monte Carlo realizations of PBH initial conditions to average over sampling noise. 

Moreover, in the low-mass PBH regime (\( M_{\rm PBH} \lesssim 10\,M_\odot \)), the gravitational influence of individual PBHs on gas collapse is subdominant compared to collective tidal effects and background halo dynamics. This suggests that any residual two-body noise does not qualitatively affect the suppression trends we report. Nevertheless, we acknowledge that future studies targeting low PBH masses would benefit from higher-resolution DM particles (i.e., \( m_{\rm BDM} \ll M_{\rm PBH} \)) or semi-analytic treatments of PBHs to fully eliminate potential artifacts in this limit. For higher PBH masses, i.e. $m_{\mathrm{PBH}} \sim 10^6 \, M_{\odot}$, a similar behavior regarding the acceleration or delay of star formation has been observed in \cite{zhangHowMassivePrimordial2025}, although the delay in that case is due to feedback mechanisms not considered in our work.

\section{Conclusions and Future Work}
\label{sec:conclusions}

In this study, we employed cosmological $N$-body simulations to investigate the impact of primordial black holes on Population III star formation. Our results reveal a complex interplay of competing physical processes: massive PBHs can dramatically accelerate structure formation by serving as gravitational seeds, leading to significantly earlier onset of Pop III star formation, while the presence of PBHs can also induce tidal disruptions that delay or suppress star formation under certain conditions.

Our simulations demonstrate that PBHs with masses around $10^4 \, M_{\odot}$ strongly enhance early star formation, potentially conflicting with observational constraints on the timing of first star formation. The most dramatic acceleration occurs when both PBH mass and abundance are large, where the Poisson effect can shift star formation to redshifts as high as $z \sim 245$—far earlier than the $z \sim 26$ predicted by standard $\Lambda$CDM models.

Conversely, lower-mass PBHs ($10 - 10^3 \, M_{\odot}$) exhibit abundance-dependent behavior, creating complex interactions within DM halos. At low abundances, these PBHs can delay star formation through tidal heating effects that counteract gas cooling, while at higher abundances they accelerate formation through gravitational enhancement of density fluctuations.

These competing effects enable us to derive meaningful constraints on the mass and abundance of PBHs contributing to the universe's dark matter content. By requiring that Pop III star formation occurs within observationally motivated redshift ranges, we exclude significant regions of PBH parameter space, particularly for massive PBHs at high abundances.

Our work establishes a new avenue for constraining primordial black holes through their effects on early star formation, complementing existing constraints from gravitational wave observations, microlensing surveys, and cosmic microwave background measurements. The derived constraints are particularly relevant for PBH masses in the range $10 - 10^4 \, M_{\odot}$.

Several promising directions emerge for future investigation. First, extending our analysis to non-monochromatic PBH mass distributions would provide more realistic constraints, as primordial formation mechanisms typically produce extended mass spectra rather than delta functions. Second, incorporating more sophisticated gas physics—including molecular hydrogen cooling, stellar feedback, and chemical enrichment—would enable more precise predictions of Pop III star formation timing and efficiency. Third, expanding the simulation volume and resolution would allow investigation of environmental effects and the statistical distribution of formation times across multiple halos.

Additionally, our framework could be extended to study PBH effects on subsequent generations of star formation and early galaxy assembly, potentially providing constraints across a broader range of cosmic epochs. The methodology developed here also provides a foundation for investigating other exotic dark matter candidates that might similarly influence early structure formation through gravitational or non-gravitational interactions.

Finally, future observations of the earliest stars and galaxies with telescopes such as the James Webb Space Telescope and 21-cm cosmology experiments will provide crucial observational anchors for refining these theoretical predictions and strengthening constraints on primordial black hole dark matter scenarios.

\acknowledgments
This work is partly supported by the U.S.\ Department of Energy grant number de-sc0010107 (SP). 

\appendix

\section{Isocurvature Perturbations and Initial Conditions}

In the present study, PBHs are injected into the zoom-in initial conditions using a machine-learned phase-space distribution derived from background dark matter particles. While this approach ensures consistency with the bulk CDM kinematics, it does not explicitly include additional isocurvature power arising from the discrete nature of PBHs, as modeled in Ref.~\cite{liuEffectsStellarmassPrimordial2022}. These small-scale isocurvature modes, which act as Poissonian fluctuations in the initial density field, have been shown to significantly impact early structure formation by seeding the collapse of minihalos at higher redshifts. For instance, Ref.~\cite{liuEffectsStellarmassPrimordial2022} demonstrates that including such perturbations can shift the collapse redshift $z_{\rm col}$ earlier by $\Delta z \sim 1$--$3$ depending on PBH mass and abundance.

The omission of explicit isocurvature power in our initial conditions may lead to a conservative underestimate of PBH-induced structure enhancement. However, the qualitative trends observed in our simulations—particularly the suppression of Pop III formation for low-mass PBHs and its enhancement for higher masses—remain robust and are dominated by nonlinear feedback and tidal effects. Incorporating PBH-induced isocurvature modes into initial conditions represents an important direction for future work to fully capture the gravitational influence of PBHs on early star formation.

Regarding the gravitational softening length for PBHs, we adopt a comoving value of $\epsilon_{\rm PBH} = 0.01\, h^{-1} \mathrm{kpc}$, in line with previous studies~\cite{liuEffectsStellarmassPrimordial2022}, and have verified that this resolution is sufficient to track halo collapse without artificial suppression due to numerical effects.


\bibliographystyle{apsrev4-1}
\bibliography{references}
\end{document}